\documentclass[a4paper,fleqn,usenatbib]{mnras}

\usepackage[T1]{fontenc}
\usepackage{ae,aecompl}

\usepackage{graphicx}	
\usepackage{amsmath}	
\usepackage{amssymb}	
\usepackage{color}
\usepackage{float}
\usepackage{ulem}

\def\eq{\begin{equation}}
\def\en{\end{equation}}

\def\gtrsim{\raisebox{-0.3ex}{\mbox{$\stackrel{>}{_\sim} \,$}}}

\def\etal{{\it et al}\thinspace}

\def\apj{{\it Ap.J.}\thinspace}
\def\apjl{{\it Ap.J. Lett.}\thinspace}
\def\apjs{{\it Ap.J. Suppl.}\thinspace}

\def\aap{{\it A\&A}\thinspace}

\def\mnras{{\it MNRAS}\thinspace}
\def\P3hat{{\mathaccent 94 P}_3}

\title{Quasi-Periodicities in the Anomalous Emission Events in Pulsars B1859+07 and B0919+06}

\author[Haley Wahl, Dan Orfeo, Joanna Rankin \& Joel Weisberg]{Haley M. Wahl,$^{1}$\thanks{E-mail: Haley.Wahl@uvm.edu} Daniel J. Orfeo,$^{1}$ Joanna M. Rankin$^{1}$ and Joel M. Weisberg$^{2}$
\\
$^{1}$Physics Department, University of Vermont, Burlington, VT 05405\\
$^{2}$Physics and Astronomy Department, Carleton College, Northfield, MN 55057}

\date{Accepted 2016 June 30. Received 2016 June 30; in original form 2016 April 27}

\pubyear{2016}

\begin{document}
\label{firstpage}
\pagerange{\pageref{firstpage}--\pageref{lastpage}}
\maketitle

\begin{abstract}

A quasi-periodicity has been identified in the strange emission shifts in pulsar B1859+07 and possibly B0919+06. These events, first investigated by Rankin, Rodriguez \& Wright in 2006, originally appeared disordered or random, but further mapping as well as Fourier analysis has revealed that they occur on a fairly regular basis of approximately 150 rotation periods in B1859+07 and perhaps some 700 in B0919+06.  The events---which we now refer to as ``swooshes"---are not the result of any known type of mode-changing, but rather we find that they are a uniquely different effect, produced by some mechanism other than any known pulse-modulation phenomenon. Given that we have yet to find another explanation for the swooshes, we have appealed to a last resort for periodicities in astrophysics:  orbital dynamics in a binary system.  Such putative ``companions'' would then have semi-major axes comparable to the light cylinder radius for both pulsars.  However, in order to resist tidal disruption their densities must be at least some 10$^5$ grams/cm$^3$---therefore white-dwarf cores or something even denser might be indicated. 

\end{abstract}




\begin{figure*}
\begin{center}
\begin{tabular}{@{}lr@{}}
{\mbox{\includegraphics[width=91mm,angle=0]{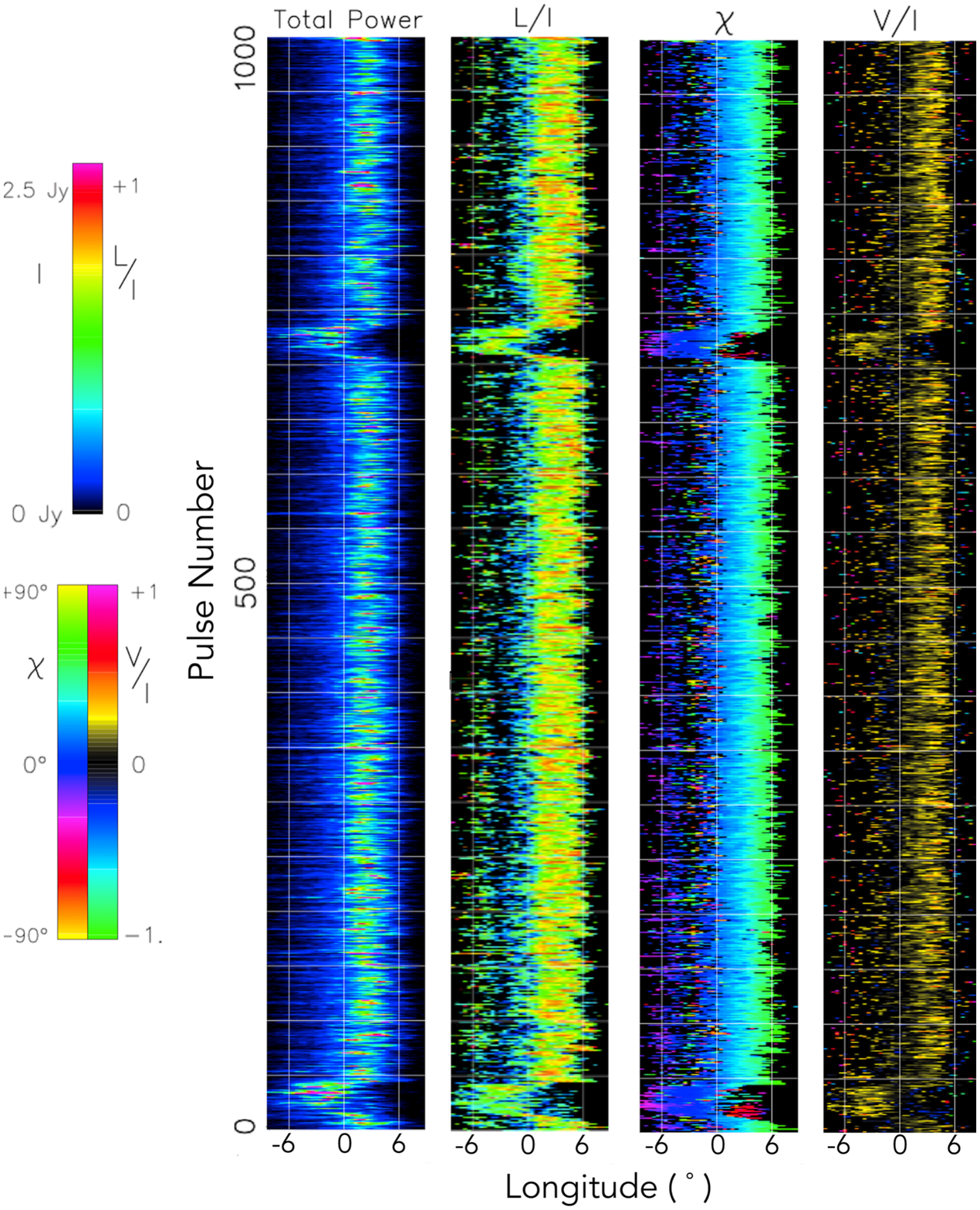}}} 
\hspace{-0.7 cm}
{\mbox{\vspace{-0.6 cm}\includegraphics[width=94.5mm,angle=0]{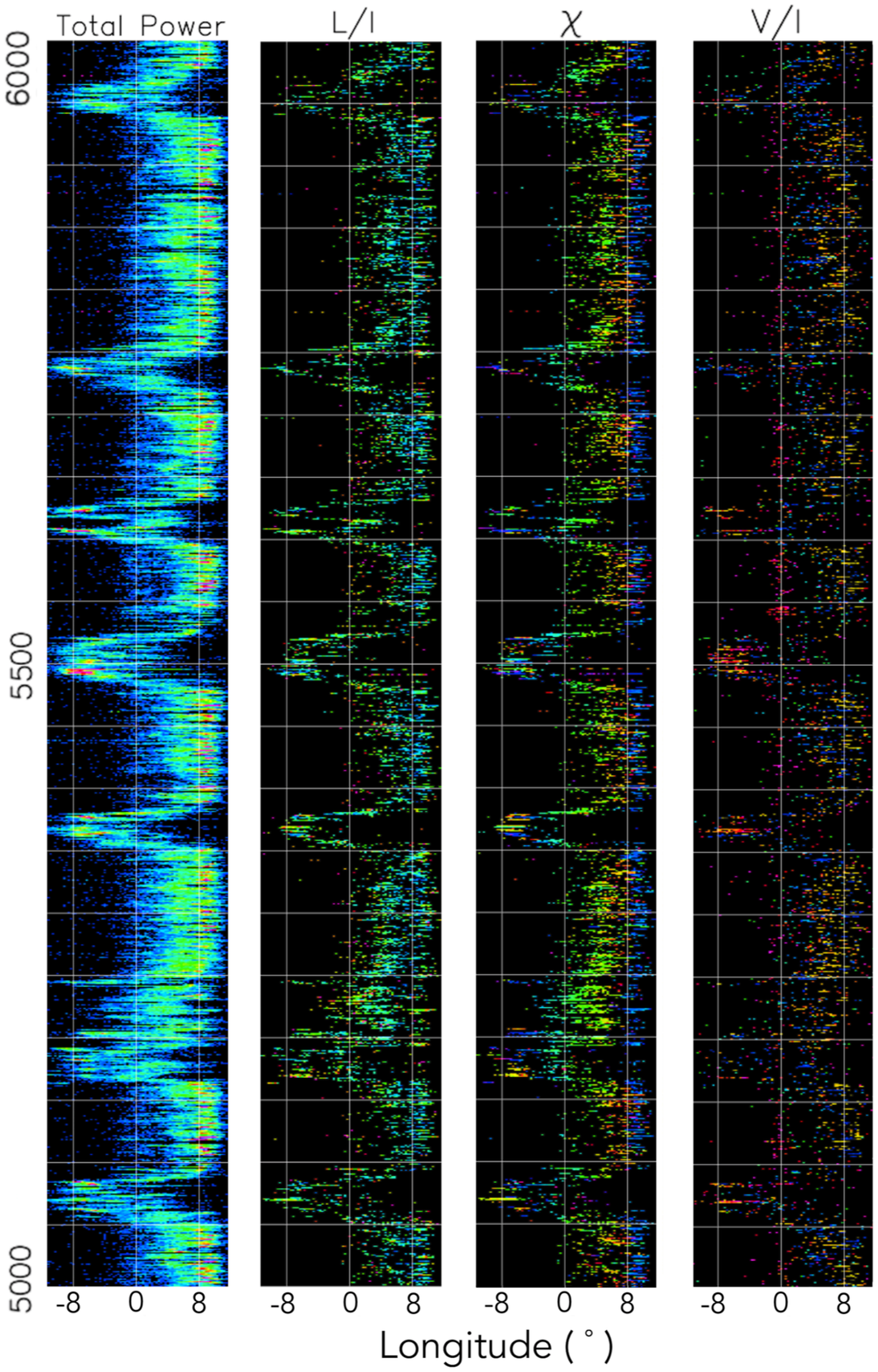}}}\\
\end{tabular}
\vspace{-0.7 cm}
\caption{Single-pulse polarization displays of the swoosh events in 1000-pulse 
intervals for both pulsars:  two swooshes can be seen in B0919+06 (left) and some 7 in 
B1859+07 (right). The total power $I$, fractional linear $L/I$, polarization position angle $\chi$, and fractional circular polarization $V/I$ are colour-coded in each of four columns according to their respective scales at the left of the diagram.}
\label{fig1}
\end{center}
\end{figure*}

\section{Introduction}
Pulsars are rapidly-rotating, highly-magnetized neutron stars that emit electromagnetic radiation along their magnetic axes which are tilted with respect to their rotation axes. This circumstance causes their radio signals to sweep across our sightline as the star rotates, producing discrete, regular pulses. Such pulse trains or sequences can be complicated by several known phenomena including \textit{nulls} (one or more consecutive pulses with no detectable power) and \textit{mode changes} (abrupt switches between two distinct emission patterns). Both of these effects have been studied extensively and are broadly consistent with our overall understanding of pulsar action--that is, emission forms cones around the magnetic axis that in turn are comprised of a rotating ``beamlet" system.

Additional anomalous events were first detected in the emission of radio pulsars B0919+06 and 
B1859+07 by Rankin, Rodriguez \& Wright (2006, hereafter RRW). In these events, a) the regular pulsed emission gradually moved earlier in longitude over a few 
pulses, b) remained early for a dozen or more pulses---largely emptying the usual emission 
window, and c) then returned over a few pulses to the usual longitude. These events attracted notice because they did not seem to be  ``classical" mode changes, which exhibit transitions within a single rotation.  

RRW examined the events in both pulsars, drawing on six full-polarization Arecibo observations (four of B0919+06 and two of B1859+07 made between 1981 and 2005).  They attempted to work out 
the two pulsars' emission geometry and examined several possible explanations for the events.  
The possibility of some new type of mode changing was examined, but was ruled out because 
of the gradual onsets and relaxations of the events. Changes in emission altitude were also examined, but the needed displacements within the magnetosphere seemed unreasonably large.  Appeals to  ``absorption" or other effects seemed ad hoc.  In short, RRW were unable to identify a plausible physical cause for these observed events.  However, they did show that both the normal and event profiles seemed to stem from partial illumination of the stars' polar caps, as artificial profiles constructed from a comparable number of normal and event pulses produced core-cone triple profiles with the usual core and conal angular dimensions (Rankin 1993, A \& B).

Several other studies have drawn attention to these two pulsars.  Mitra \& Rankin (2011) conducted sensitive single pulse polarimetry on most of the pulsars originally categorized by Lyne \& 
Manchester (1988) as ``partial cones"---including B0919+06---and their analysis tends to support 
the interpretation of RRW to the effect that the events entail a partial illumination of the pulsar's 
emission cone, late for the normal emission and earlier during the events.  Evidence for long 
term periodicities on the order of tens of years in the timing of B0919+06 were developed by Lyne \etal\ (2010) and 
Shabanova (2010) and thoroughly studied recently by Perera \etal\ (2015).  A similar study on 
B1859+07, again drawing on the unexcelled pulsar timing archive at the Jodrell Bank Observatory, has been carried out by Perera \etal\ (2016). Most recently, Han \etal\ (2016) published the results of $\sim30$ hours of B0919+06 observations using the Jiamusi 66-m Telescope and found events similar to those identified by RRW.

In this paper we undertake a new study of these emission events in pulsars B0919+06 and 
B1859+07.  In order to emphasize their distinct character, we refer to the events as ``swooshes," 
returning to the (unpublished) terminology used by RRW in the preparation of their paper. First noticed in the 1990s in Arecibo observations of B0919+06, the phenomenon was ignored until it was again seen by RRW---and now in this work with three different Arecibo instruments. That the events were also seen with the Jiamusi Telescope rules out any instrumental cause. After
initial analysis of these two pulsars by RRW, it had become clear that an entirely new approach was needed to investigate and explain the swoosh phenomenon, and we began by carrying out much longer and more extensive Arecibo observations of both pulsars.    Our analyses produced evidence of a perplexing periodicity in both stars, which in turn opened up new potential avenues of interpretation.  \S 2 describes the observations. \S 3 and \S 4 examine events in each star in more detail, and \S 5 and \S 6 discuss the evidence for their periodicity. \S 7,  \S 8, and  \S 9 then discuss possible mechanisms for the swooshes. Finally \S 10 summarizes the results and discussion.

\begin{figure}
\begin{center}
\hbox{\hspace{-0.55 cm}\includegraphics[width=90mm, angle=0.]{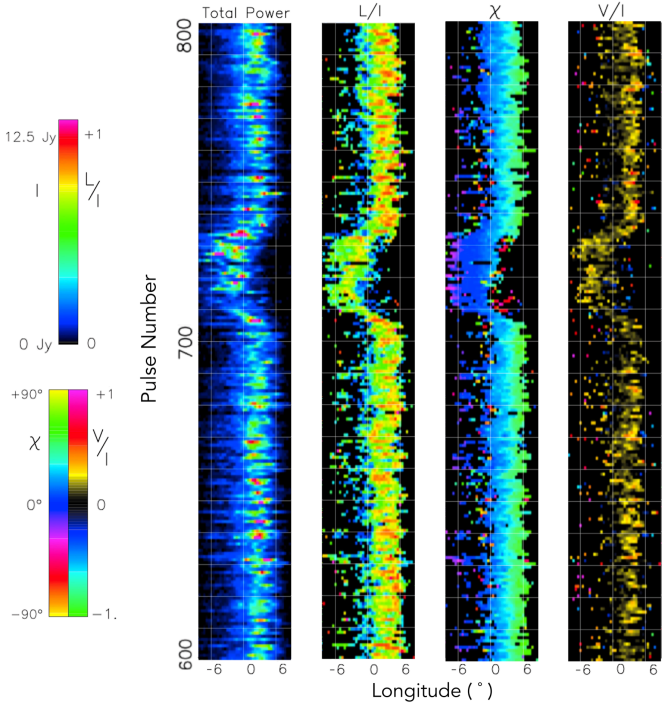}}
\vspace{-1.1 cm}
\caption{Higher resolution 200-pulse display showing the second B0919+06 swoosh in 
Figure~\ref{fig1}.}
\label{fig2}
\end{center}
\end{figure}

\begin{table}
\begin{center}
\caption{Observational parameters}
\begin{tabular}{cccccc}
\hline
\hline
Pulsar & Date & MJD &  Length & Number  \\
 RF (MHz) & (m/d/yr) & & (pulses) &  of Events \\
 \hline
{\
 \bf B0919+06} \\
1425  &   8/03/2003      & 52854 & 1115 & 2\\
327  &   10/04/2003      & 52916 & 4180 & 1\\
325.85  & 5/05/2013 & 56417 & 17259 & 6\\
325.85  & 10/11/2013 & 56576 & 17810 & 7\\
1392.5   & 10/12/2013 & 56577 & 13994 & 4\\
1494  & 8/23/2015 & 57257 & 16558 & 6\\
\\
2250 & 4/18/2015 & 57130 & 2088 & 1\\
2250 & 7/12/2015 & 57215 & 53053 & 22\\
2250 & 7/14/2015 & 57217 & 67611 & 21\\
2250 & 8/17/2015 & 57251 & 65003 & 26\\
2250 & 8/18/2015 & 57252 & 35306 & 16\\
2250 & 8/19/2015 & 57253 & 25001 & 6\\

  \\
{\bf B1859+07} \\
1525  & 04/10/2003 & 52739 & 1021 & $\sim$7\\
1520  &  1/2/2005   & 53372 & 2096 & $\sim$12\\
1392.5 & 3/26/2013 & 56377 & 6460 & $\sim$40\\
1392.5 & 4/8/2015 & 57121 & 12195 & $\sim$171\\
1394 & 12/5/2015 & 57361 & 12154 & $\sim$141\\
 \hline
 \label{Table1}
 \end{tabular}
 \end{center}
 \normalsize
 \end{table}

\section{Observations}
These new observations were carried out using the upgraded Arecibo Telescope 
in Puerto Rico with its Gregorian feed system, 327-MHz (``P band") or 1100-1700-MHz 
(``L band") receivers, with either Wideband Arecibo Pulsar Processors (WAPPs) or Mock 
spectrometer backends.  At P band four 12.5-MHz bands were used across the 50 MHz available.  
Four nominally 100-MHz bands centered at 1170, 1420, 1520  and 1620 MHz were used 
at L band, and the lower three were usually free enough of RFI such that they could be 
added together to give about 300-MHz bandwidth nominally at 1400 MHz.  The four 
Stokes parameters were calibrated from the auto- and cross-voltage correlations computed by the spectrometers, corrected for interstellar Faraday 
rotation, various instrumental polarization effects, and dispersion.  The resolutions of 
the observations are all about a milliperiod. The Gregorian and modified Julian dates 
as well as the lengths of the observations are listed in Table 1.

\section{Characteristics of the Swoosh Events}
The primary signature of the swooshes in pulsars B0919+06 and B1859+07 is a shift---to 
an earlier longitude---at which the emission occurs.  As seen in Figure~\ref{fig1}, the shifts 
during these swooshes appear nearly identical in B0919+06 and B1859+07 when examined 
on an interval of 1000 or more pulses.  However, some differences are perceptible between the events in each pulsar when  they are examined in closer detail.

Fig. ~\ref{fig1} exhibits key features of the swooshes in the two pulsars:  one can see 
that the swoosh transitions are always in the earlier direction and that both onset and termination are gradual over a few pulses.  Also, the polarized character of the emission remains very similar during a swoosh. $L/I$ and $V/I$ show no systematic change, and $\chi$ rotates linearly even while the peak emission moves to earlier longitudes. Finally, there is generally little or no change in the intensity of the individual pulses during a swoosh, but missing pulses are seen in B0919+06, see details below, which is why we choose to call these events ``swooshes" rather than ``flares."

\subsection{B0919+06}
Our analysis of B0919+06 focuses on twelve observations detailed in Table~\ref{Table1}. 
Half of the pulse sequences are taken from our own Arecibo observations while the 
other half was acquired by  Han \etal\ (2016). Using the 1000-pulse displays, (see Fig.~\ref{fig1}) 
these events stand out as -10\degr\ shifts in the longitude of the emission window that persist for anywhere from 32-70 pulses. Though the length and extent of the shifts can vary, B0919+06 events follow a consistent pattern: First, weaker emission is seen for 2-5 pulses which can sometimes be so drastic as to appear as a null.  Second, during the maximum longitude shift of the next several dozen pulses, the usual emission region largely empties, the polarization angle exhibits a modest progressive rotation of up to about --18\degr\ ,  and $L/I$ decreases from some 80\% to 50\%.  Finally, the emission gradually returns to its usual longitude over a few pulses, sometimes followed by another weakening or apparent nulling for 2-5 pulses.  This pattern is consistent for each B0919+06 event throughout all of our observations, following the same pattern  over the more than three decades of Arecibo observations at all frequencies. Though the signatures of each events are almost identical on a gross scale, Han \etal\ were able to classify them into four subtly different categories (see \S 4 for details).

\subsection{B1859+07}
The B1859+07 events  appear almost identical to those of B0919+06 when 
viewed on the 1000-pulse display of Fig.~\ref{fig1}. However, upon closer examination, the B1859+07 swooshes are seen to have much more varied and irregular patterns: their duration can be anywhere from 20-120 pulses and they appear to have different morphological characteristics (see \S 4). They are also not always fully distinct and separated from each other, which can make it difficult to identify individual events. This circumstance is the origin of the uncertainty in the number of B1859+07 events listed in Table 1. These intervals of ``connected" events appear a few times in the early shorter observations taken over the course of about ten years but increase in frequency in the later MJD 56377 observation. The even-more-recent April 2015 observation (MJD 57121) shows a further increase in the frequency of these irregular, consecutive events, and events in general, to the point where the number of swooshes in each 1000-pulse interval almost doubled compared to relatively consistent tallies in the previous observations. A long December 2015 observation also showed an increase in the frequency of these events compared to observations from the previous decade, though the event frequency is slightly below that of the April 2015 observation. Further investigation with longer observations is needed to examine the cause of this sudden increase in event frequency.

\section{Do Each of the Swooshes Look the Same?}

In RRW, all swooshes studied in both B0919+06 and B1859+07 seemed to exhibit one specific pattern in which the emission gradually decreased in longitude, stayed in that position for a number of pulses, and made a gradual transition back to the normal emission longitude. RRW concluded that the events in both stars followed this pattern, but upon further analysis on longer observations, it was discovered that the events in B0919+06 as well as B1859+07 can come in many different shapes and sizes. Han \etal\ were able to categorize the events in B0919+06 into four categories: \textit{$\Pi$ Type} (where the emission abruptly jumps
to an earlier longitude, stays there for a few tens of periods, and then abruptly jumps back down to its normal longitude), \textit{M Type} (where the emission moves to an earlier longitude for some 20 periods and then gradually returns back) (see Fig.~\ref{fig2} for an example), \textit{$\Lambda$ Type} (where the emission sharply decreases in longitude and then sharply transitions back without remaining in the early state for more than 1-5 pulses), and \textit{ $\lambda$ Type} (where the emission moves sharply to an earlier longitude and then slowly transitions back to the regular state). 

Just as Han \etal\ were able to categorize the shapes of events seen in B0919+06, we were able to perform similar analyses on the B1859+07 swooshes and sort them into seven categories based on their morphology. Due to the fact that the B1859+07 swooshes are very similar to those of B0919+06 when analyzed in detail, we take from the Han \etal\ analysis the first four morphological categories of B1859+07 swooshes: $\Pi$, M, $\Lambda$, and $\lambda$ Types. The last three B1859+07 swoosh categories are patterns which do not appear in any current B0919+06 observations: two brief consecutive shifts to an earlier longitude which essentially appear as two connected $\Lambda$ Type events with no more than 1-5 pulses between them (where the emission may or may not return to its original longitude in the middle), we refer to as \textit{$\omega$ Type}; events which feature brief sharp transitions to an earlier longitude like a $\Lambda$ Type except fill the window from starting longitude to ending longitude during the event, we refer to as \textit{$\Delta$ Type}; and finally the \textit{O Type,} which are similar to the $\Delta$ Type but have slower transitions causing the entire event to appear rounded.

\begin{figure}
\begin{center}
\hbox{\hspace{-.5 cm}\includegraphics[width=90mm,angle=0.]{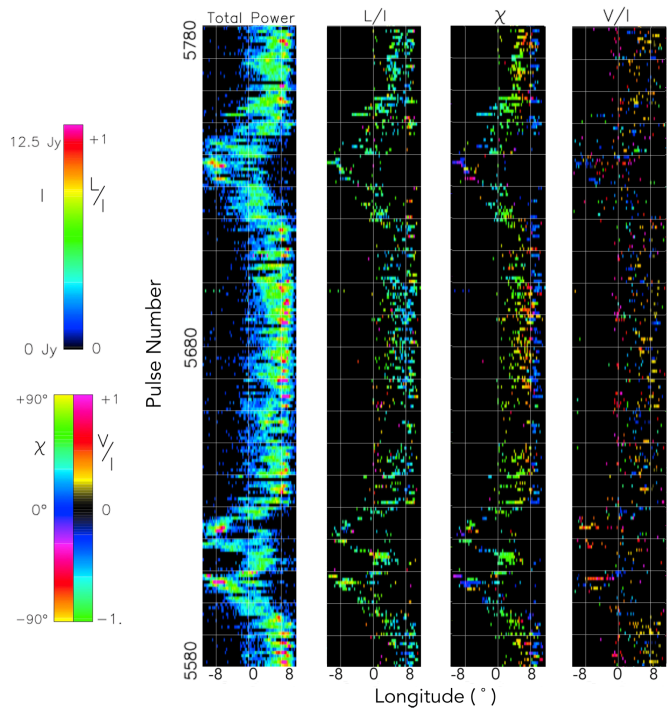}}
\vspace{-0.9 cm}
\caption{Higher resolution 200-pulse display of the fifth and sixth major B1859+07 
swooshes in Figure~\ref{fig1}.}
\label{fig3}
\end{center}
\end{figure}

Figure~\ref{fig3} gives a 200-pulse display showing two of these events in MJD 56377: an $\omega$ Type and an M Type event.  Of additional note in this observation are several long intervals in which no events occur, long stretches of $\omega$ Type events, single-pulse nulls, and longer nulls lasting 2-10 pulses. Analysis of the shorter observations (MJD 52738 and 53372) reveals additional  irregular behavior. The fact that the categorizations in both our analysis and the Han \etal\ analysis seem very similar and the types of events look very much the same when studied closely at small scales strongly indicates that a similar phenomenon is causing the events in both pulsars and therefore allows them to be studied side-by-side.

\vspace{1 cm}

\begin{figure} 
\hbox{\hspace{-2.2 cm}\includegraphics[width=130mm, height=170mm, angle=0.]{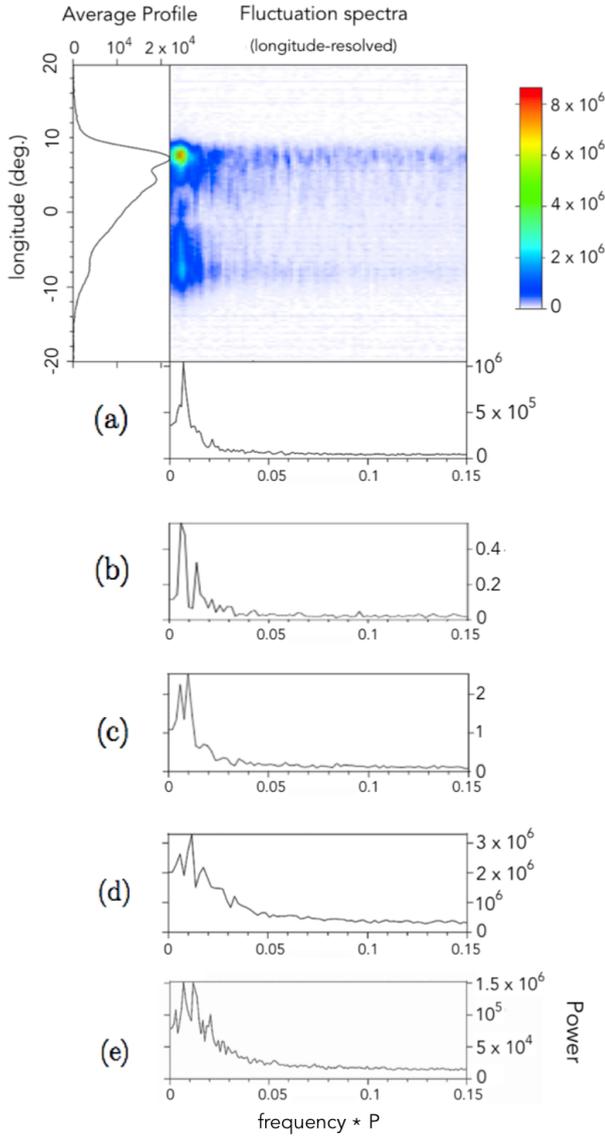}}
\vspace{-0.4 cm}
\caption{Fluctuation spectra of B1859+07, showing evidence for event periodicity. (a) The display for  MJD 56377 contains three contiguous panels, with the pulse profile on the left, the longitude-resolved fluctuation spectrum in the 
top right, and the overall fluctuation spectrum at the bottom; while the overall fluctuation spectra for
 MJD  52739, 53372, 57121, and 57361 are shown in (b) through (e), respectively. The profile and longitude-resolved spectra are shown in the  left panels and the total fluctuation spectrum in the lower one. Note that all of the overall fluctuation spectra are plotted only to 0.15 cycles/period. The feature at 0.00685 cycles/period corresponds to a periodicity of 146 rotation periods. Fourier transforms of length 1024 were used for the MJD 56377 analysis and 512 for the others. The different amplitude scales  result from use of the two different backend processors.}
\label{fig4} 
\end{figure}  

\section{Identification of a Complex Swoosh Periodicity in Pulsar B1859+07}

Though the events in B0919+06 and B1859+07 appear to be somewhat different when studied 
in detail, they both exhibit a similar gradual early shift pattern of the emission longitude. 
As seen in Table~\ref{Table1}, the events seem to appear much more frequently in B1859+07 than they do in B0919+06, which prompted further analysis of the former star.  Moreover, the 
B1859+07 swooshes give the impression of occurring with a rough regularity of 
several hundred pulses as we see in Fig.~\ref{fig1}.

We therefore first studied the MJD 56377 observation of B1859+07 by computing a longitude-resolved fluctuation spectrum, so as to investigate the reality of any event periodicity. This spectrum, shown in the top panel of Figure~\ref{fig4}(a), remarkably
displays a well defined periodic feature with a peak frequency at 0.00685 cycles/period 
($P$=146 pulsar rotation periods) associated most strongly with the leading and trailing edges of the profile.  The latter property ties this principal fluctuation to the events 
since a longitudinal shift of emission would be most evident at these longitudes. A more accurate value for the total fluctuation frequency can be computed by using all of the spectral components with intensities greater than the half-power point ($\Delta f$ is this half-power 
width), giving $P$ =153$\pm$3 periods. Note, however, that the ``Q" [=$f/\Delta f$] of the feature is not very large, perhaps 2, so the effective error is much larger than that given 
just above.  This squares with our further efforts to measure the period of the swooshes with a periodigram analysis, which showed little regularity in the intervals between them.  

Indeed, it is difficult to understand how the strong feature in the MJD 56377 fluctuation 
spectrum arises when examining the distribution of intervals between swooshes.  The 
majority of the intervals fall between 50 and 100 periods.  However, there are several 
long intervals without swooshes, one 280 and another 351 periods.  

Nonetheless, given the identification of this fairly regular modulation in our MJD 56377 
observation, it seemed important to see what our other observations showed.  The total 
fluctuation spectra of the MJD 52739, 53372, 57121, and 57361 observations are shown 
in Fig.~\ref{fig4}(b)-(e), respectively; and given their shorter length or more complex 
modulation, FFTs (Fast Fourier Transforms) of length 512 were used for them, while a length of
1024 was employed for the results of Fig.~\ref{fig4}(a).  The short MJD 52739 observation 
of Fig.~\ref{fig4}(b) shows a strong feature at 0.0068 cycles/period with a period of 148$\pm$5 stellar rotations.  
There is also power at 0.014 cycles/period or 73$\pm$3 periods, which seems to be 
a harmonic of the main feature.  The somewhat longer MJD 53372 spectrum of Fig.~\ref{fig4}(c) 
has two strong peaks at 170 and 102 periods; however, when 
all of the fluctuation power in the first seven components is weighted and averaged, 
the result is 152$\pm$8 periods, again with a small $Q$. This all said, 
we reemphasize that the feature widths are all much larger than the errors in the spectral 
peaks given just above, or equivalently their `Q's are again of order 2.  Therefore, it 
seems that the swooshes have a consistent average period while exhibiting substantial 
irregularities in their intervals of occurrence.

The total fluctuation spectra of the MJD 57121 and 57361 observations (Figs.~\ref{fig4}(d)-(e)) show a somewhat different spectrum.  This was not too surprising given the 
above discussion of their more frequent and difficult to interpret swooshes.  Both are 
also the longest of our observations at more than 12,000 pulses.  Fluctuation power in 
these observations is spread over a wider range than in the other observations.  Both 
spectra show two peaks.  In the 57121 spectrum one is at 170 and the other at 85 periods; 
whereas the 57361 has peaks at 143 and 81 periods, all with errors of  some 10 periods.  
A weighted average of the 57121 spectral power up to the sharp decrease at about 
0.014 cycles/period, gives a period of some 144$\pm$10 rotation periods.  For the 
57361 spectrum the errors are such that the two features could be harmonically 
related as indicated above, though all the $Q$s are small.

\section{Are the swooshes in B0919+06 Quasi-Periodic as Well?} 

The swooshes in B0919+06 are much less frequent than they are in B1859+07. In our six observations totaling some 8.5 hours and 70,916 pulses, we have identified only 26 swoosh events, while Han \etal\ were able to identify a total of 92 abnormal emission events in some 30 hours or 248,062 pulses total. 

The three shortest intervals between swooshes in the full set of B0919+06 observations are 
292, 495 and 549 pulses.  The next largest group are 728, 779, 806, and 880.  
All the others are much longer, the 4180-pulse MJD 52916 sequence 
shows but a single swoosh, and in two instances the interval is longer than 
10,000 pulses.  Previous observations taken with  the Arecibo Telescope showed 
shortest intervals between swooshes as 662, 837, and 900 pulses with the next largest 
group being 1508, 1742, 1924, 1529, and 1341 pulses. 

A rough approximate period using just our Arecibo observations was calculated to be about 700 pulses or 300 seconds.  With the introduction of the Han \etal\ sequences, the 
B0919+06 events could be studied over much longer time scales, which revealed that their 
occurrence is at least as complex as in B1859+07. For B1859+07, a consistent long term periodicity 
of some 150 periods is found using Fourier techniques, whereas mapping of the intervals 
between events shows many smaller and larger intervals. We have investigated the Han \etal\ B0919+06 sequences using Fourier methods as well, and we find no indication of periodicity at all, only low frequency fluctuation power at periods greater than several hundreds stellar 
rotations. 
		
\section{How Can We Interpret the swoosh Phenomenon?}
The putative periodicity in the swoosh events in pulsar B1859+07, introduces a 
variety of new directions and complications into any effort to interpret the swooshes 
physically.  Let us summarize---
\begin{itemize}
	\item the swooshes are not conventional mode changes for a variety of reasons.  
		They typically begin and end gradually.  No other pulsar switches modes by simply 	
		altering the longitude range of its emission [see Rankin (1986) for 
		references to studies of the then known mode switching pulsars].
	\item the swooshes always represent intervals of earlier emission.  RRW's 
		suggestion of ``partial conal" emission remains possible, but we have no 
		physics to understand why the illumination would usually be late but then  
		earlier during the swooshes [See Mitra \& Rankin (2011) for a full study 
		of ``partial cone" emission].
	\item that the swooshes are caused by an emission-height change remains possible, but we 		think unlikely. In no pulsar has this effect been clearly demonstrated or a physical cause 		identified. Ad hoc assumptions are 
		needed to estimate such a change and again, why such a height change should 			occur (e.g. why the height should increase rather than decrease).
	\item no emission mechanisms are known that would produce quasi-periodic 
		shifts of the sort observed in the swooshes.  Some other emission effects 
		are known with very long time scales over multi-hour long time scales (e.g., the 			carousel rotation slowdown after B-mode onset in pulsar B0943+10; see 					Rankin \& Suleymanova 2006), but no periodicity has been identified with effects such 		as these.
	\item pulsations, a form of surface oscillations known to exist in white dwarfs, have been 			 explored as a possible cause for periodic phenomena in pulsars (e.g., Rosen \& 			 Clemens 2008). In terms of the events in our pulsars, we do not believe that 				 oscillations are the cause for three reasons: 1) there is no conclusive evidence that 		 	 these oscillations/pulsations exist in pulsars; 2) the expected $\sim$ms oscillation 		          timescales  would be much smaller than the $\gtrsim$1 s time scales of our events; 		 and 3) such oscillations tend not to have a constant period but quicken as they lose 			 energy, which is not seen in our swooshes.
\end{itemize}

\section{Could the swooshes Reflect an Orbiting Body or Bodies?}
Let us consider an alternative explanation. 
In astrophysics, a fundamental physical cause of periodicity is orbital dynamics.  
However, ---
\begin{itemize}
	\item periodicities due to orbital causes tend of have a high Q so it is difficult to  
		understand how a ``companion" orbiting the pulsar could cause the very rough 
		periodicities we apparently see in B1859+07 and B0919+06.  
	\item the time displacements are not easy to understand:  the swooshes move 
		the emission earlier in both pulsars.  A highly eccentric orbit might produce 
		a displacement  in one direction for one part of the orbital period, but it is likely
		that secular variations of the orbital elements would eventually produce shifts in 
		both directions.
	\item binary orbits are not known for such short periods: 153 rotation periods for 
		B1859+07 would be only 99 secs, and 700 rotations for B0919+06 only 
		300 secs.  
\end{itemize}
These conclusions, however, strictly follow only if the orbit is a simple binary.  More complex 
orbital effects follow from more complex systems, including the possibility of a chaotic 
system.  Such close orbital mechanics would certainly entail decay by gravitational 
radiation, so the finite lifetime of the system must be considered.    
	
\section{What if the Swooshes did Reflect an Orbiting Body or Bodies?}

Given that we cannot explain the swooshes and their putative periodicities by other 
means, we here explore what kind of orbital systems might produce the effects we 
observe. Specifically, we describe the consequences of interpreting the inter-swoosh quasi-period $P_s$  as the orbital period $P_b$ of such systems.

If the companion had a mass much smaller than 
that of  the neutron star, we could use Kepler's Third Law $(M/{\rm M_{\sun}}) (P/{\rm yr})^2 =
 (a_c/{\rm AU})^3$  
to compute the companion's orbital semi-major axis $a_c$, 
given a system mass $M\sim1.4\ {\rm M}_{\sun}$ and orbital period $P$.
For PSR B1859+07, the orbital period $P=P_s$ is $\sim$150 secs or 2.1 $\mu$y, so that $a_c$ 
$\sim$180 $\mu$AU
or some 27,000 km.  Remarkably, this is only a little larger than the pulsar's light-cylinder radius 
of 20,560 km! If we do a similar calculation for B0919+06 using the period of about 700
pulses, we find an $a_c$ value of 760
 $\mu$AU or some 120,000 km, as compared to a light cylinder of radius 30,560 km. 

These estimates show that the semi-major axes of the orbits are larger, but only a little 
larger, than the light-cylinder radius for both pulsars.  Were the orbits elliptical, 
the ``satellite" could pass close to or even inside the light cylinder.  It is also interesting that 
the closer orbit to its light cylinder is that of B1859+07, where we apparently see more of the swooshes than for B0919+06. Given that the 
\textit{orbital} period of the ``satellite" and the \textit{rotation}
period of the pulsar are asynchronous, one can further 
 speculate that a variety of effects might be produced, as a function of  the (unknown) orientation of the ``satellite" with respect to our sightline when it passed close to the pulsar.  

The lack of cyclical emission-longitude variations having periodicity $P_s$ (aside from the low duty-cycle
swooshes themselves) indeed  indicates that the companion mass must be relatively low,
thereby validating the assumptions underlying the above calculations. Even if the swoosh
longitude offsets themselves are interpreted as the pulsar reflex motion, their $\sim$10 ms
amplitudes indicate a projected pulsar semi-major axis $a_p \sin i$ of only $ \sim$10 light-ms,
which is much less than the characteristic size of the {\it companion} orbit, as calculated
above.

Could any companion persist in such an environment so close to the pulsar?  Tidal disruption would be a major issue, and the ability of a ``satellite" to resist disruption provides a constraint on its density.  Disruption occurs when $M/a_c^3$ > $\rho_{\rm c}$, where $\rho_{\rm c}$ is the companion's density (Harwitt 1973). For B1859+07 this limit on the density is about $10^5$ g/cm$^3$, suggesting that only a body having a density of a white dwarf core or greater could so persist.  Indeed, pulsar-white dwarf binaries  
are known, but with much longer periods; however, one such system with 
a period of two hours has been identified (Bailes \etal\ 2011), while white dwarf-white dwarf binaries with orbital periods as short as 12.75 min have been discovered (Brown \etal\ 2011). Furthermore, the lack of observed binary systems of shorter known periods could easily be due to the difficulty of detecting them.

\section{Summary}

We report extensive new Arecibo observations and analyses studying the strange 
swooshes in pulsars B0919+06 and B1859+07.  We find much more complexity 
that was earlier reported by Rankin \etal\ (2006); however, we confirm the basic effect involving usually gradual, temporary, about 10\degr\ shifts in the emission window to earlier longitudes.  Here also, we provide some evidence for a quasi-periodicity in pulsar B1859+07 and perhaps in B0919+06.  We examine a number 
of effects that might produce the swooshes and find no obvious mechanism.  In 
particular, the effects might represent mode changes, but no example of a gradual 
mode change is known, so swooshes appear to be some other phenomenon.  
These difficulties have led us to explore whether some type of co-orbiting companion 
with an unprecedentedly short period and separation might be involved.  Such a 
system would entail so many effects of tide and heating that it is difficult to predict 
what the consequences might be.  Given that WD-NS and WD-WD binaries are 
known with periods 10-100 times larger one can ask what happens as those systems evolve.

\section{Acknowledgements}

HMW gratefully acknowledges a Sikora Summer Research Fellowship. Much of the work was made possible by support from the US National Science Foundation grant 09-68296 and from a Vermont Space Grant. JMW acknowledges support from NSF Grant AST-1312843. We thank Han \etal\ 2016 for access to their total-power data on PSR B0919+06.
The Arecibo Observatory is operated by SRI International under a cooperative agreement with the National Science Foundation (AST-1100968), and in alliance with Ana G. Mendez-Universidad Metropolitana, and the Universities Space Research Association. 






\bsp	
\label{lastpage}
\end{document}